\documentclass[aps,pre,twocolumn,superscriptaddress,showpacs,a4paper]{revtex4-1}
\usepackage{graphicx}
\usepackage{amssymb,amsmath}
\usepackage{wasysym}
\usepackage{textcomp}
\usepackage{setspace}
\usepackage[dvips,usenames]{color}

\def\r{{\mathbf{r}}}
\def\ro{{\mathbf{r}_0}}

\begin{document}

\title{Limits on the performance of Infotaxis under inaccurate modelling of the environment}

\author{Juan Duque Rodr\'iguez}
\email{jrduque@ucm.es}
\affiliation{Laboratory of Physical Properties TAGRALIA,
Technical University of Madrid, 28040 Madrid, Spain}

\author{David G\'omez-Ullate}
\email{dgomezu@ucm.es}
\affiliation{Department of Theoretical Physics II, Complutense
  University of Madrid, 28040 Madrid, Spain}
\affiliation{Instituto de Ciencias Matem\'aticas (CSIC-UAM-UC3M-UCM),  C/ Nicolas Cabrera 15, 28049 Madrid, Spain.}  

\author{Carlos Mej\'{\i}a-Monasterio}
\email{carlos.mejia@upm.es}
\affiliation{Laboratory of Physical Properties TAGRALIA,
Technical University of Madrid, 28040 Madrid, Spain}

\date{\today}

\begin{abstract}
  We study  the performance of  infotaxis search strategy  measured by
  the  rate of  success and  mean search  time, under  changes  in the
  environment parameters such as diffusivity, rate of emission or wind
  velocity.  We also investigate the  drop of performance caused by an
  innacurate  modelling of  the environment.   Our findings  show that
  infotaxis remains  robust as long  as the estimated  parameters fall
  within a  certain range  around their true  values, but  the success
  rate  quickly  drops making  infotaxis  no  longer  feasible if  the
  searcher  agent severely  underestimates or  overestimates  the real
  environment  parameters.   This  study  places some  limits  on  the
  performance of infotaxis, and thus it has practical consequences for
  the  design of  infotaxis  based  machines to  track  and detect  an
  emitting source of chemicals or volatile substances.
  \end{abstract}

\pacs{02.50.-r, 05.40.-a, 87.19.lt}

\maketitle

\section{Introduction}
\label{sec:introduction}

Infotaxis  is  an  olfactory  search  strategy  proposed  in  2007  by
Vergassola, Villermaux  and Shraiman \cite{Vergassola}  to address the
problem of finding  the source of a volatile  substance transported in
the environment under  turbulent or noisy conditions.  In  the lack of
such complications,  chemotaxis, i.e. moving  upwards in concentration
gradient, performs well as a search strategy and many living organisms
are  known  to use  this  strategy  to  perform their  natural  tasks.
However, when  detections are scarce  or the concentration  profile is
not smooth, it is no longer possible to estimate the concentration and
its gradient  at a  given point.  In  this regime,  chemotaxis becomes
unfeasible and  infotaxis reveals its true  significance. Some insects
are known  to navigate and  find their targets under  these scenarios,
\cite{Murlis1992,Dusen1992,Hansson1999}.     Learning    from    their
strategies   has  inspired   robotic  devices   designed   to  perform
complicated  search  tasks  with technological  applications  (finding
dangerous  substances  such  as   drugs  or  explosives  or  exploring
inhospitable                                              environments)
\cite{Russell,Mouraud2010,Martinez2006,Li2006a,Pang2006},   for  which
robustness  and   performance  of  the  search  is   of  main  concern
\cite{mejia2011,duque2014}.

Turbulent or noisy environments are usually modeled by stochastic
processes. In the simplest  model, spatio-temporal correlations in the
concentration profile  are neglected, and the number  of detections is
modeled by considering  a Poisson process at each  point of space. The
rate of detections at each point depends on the position of the source
and the parameters  of the transport process, and  is usually obtained
from the solution of an advection-diffusion equation.

The searcher agent has  a built-in model of the environment, and  it is able to
calculate the estimated number  of detections at its current location,
given  the  position of  the  source.   Instead  of knowing  the  true
position of the source, the agent uses a probability distribution that
expresses his  belief about  the position of  the source.  This belief
function is constantly updated  following Bayesian inference using the
built-in model and the number of detections actually registered by the
sensors at a given point.  The most innovative feature of infotaxis is
the criterion for  the motion of the agent:  instead of moving towards
the  most probable  position of  the source,  the agent  moves  in the
direction  where  it  expects  to  gain  more  information  about  its
position. In a  sense, it is a greedy  search in \textit{information},
as opposed to \textit{physical} space.

Infotactic searches  involving fleets of cooperative  agents have been
considered in \cite{Masson}. Extensions of the algorithm to continuous
space  and  time  and  to   three  dimensions  have  been  treated  in
\cite{Barbieri}.  Recently, Masson has  proposed an  information based
search strategy  similar to infotaxis  where the searching  agent does
not have a global space perception, \cite{MassonPNAS}.

In a  previous work, \cite{duque2014}  we analyzed the  performance of
infotaxis as the initial position  of the agent relative to the source
and  the boundary  of the  search domain  was changed.  The surprising
result  was that the  mean search  time was  not always  an increasing
function  of the  distance  to  the source:  in  some cases,  starting
further away from the source  led to shorter and more efficient search
processes. This a priori  counterintuitive result was explained by the
fact that the first step in an infotactic search is not stochastic but
deterministic, and depends only on the boundaries of the search domain
and  the parameters  of the  transport process  (rate of  emission and
correlation  length), not  on  the  position of  the  source. This  is
natural,  since  at the  beginning  of the  search  the  agent has  no
information  about the  position  of the  source,  the initial  belief
function  is uniform  and entropy  is maximum.  The search  domain was
shown to be partitioned into regions of constant first step, and these
regions are limited by smooth curves.

In this  work we extend our  study of the performance  of infotaxis to
consider two different situations:
\begin{enumerate}
\item[i)] variation  in performance as a function  of the environment,
  assuming perfect knowledge of the environment parameters.
\item[ii)] variation  in performance due  to an imperfect  modeling of
  the environment.
\end{enumerate}

In the first  case we shall assume that the  environment model used by
the infotactic agent  to do Bayesian inference is  exact, but we shall
probe infotaxis under different ranges  of values of the parameters of
the  environment.  In  the second  case we  will explore  the  drop in
performance  caused  by  an  imperfect modeling  of  the  environment,
i.e. when there is a  mismatch between the true environment parameters
and those  in use by  the agent. Both  of these problems are  of great
practical relevance:  it is essential  to know the range  of parameter
values in  which infotaxis remains  an efficient search  strategy, and
likewise  it  is  important  to  know  how  much  uncertainty  in  the
estimation or  measurement of the parameters of  the transport process
can  be allowed.   While some  of  these questions  have been  briefly
addressed  in  the recent  literature  \cite{Masson},  a thorough  and
systematic analysis as the one performed in this work was absent.

It should  be stressed  at this point  that our implementation  of the
infotaxis algorithm  includes one  differential feature from  the ones
considered  in the  literature.  In previous  studies a  \textit{first
  passage} criterion  was typically  used, i.e. the  search terminates
when the position  of the agent coincides for the  first time with the
position of  the source.  Instead, we have  used a  \textit{first hit}
criterion: the search terminates when  the entropy falls below a given
threshold,  i.e. when  the agent  has sufficient  certainty  about the
position  of  the  source.  The   reason  to  use  this  criterion  is
twofold.  First, the  agent needs  no external  information  about the
source:  it  decides  to  halt  based  on  its  own  computations  and
measurements.   Second,   it   allows   detection   at   a   distance,
i.e. successful  searches when  the agent knows  where the  source is,
even  if it  is a  distance away  from it.  This  criterion emphasizes
vicinity in the information rather  than the spatial sense.  Note that
with our criterion it could happen that the agent passes on top of the
source without actually knowing it,  and the search would continue. In
practice, however, it usually happens that when the agent first passes
by the  source, it decides not  to move and  entropy rapidly decreases
below the  threshold signaling the  source detection. In  some extreme
cases as  those studied  in this work,  deviations from  this standard
behavior could happen.

In order to assess the performance of infotaxis as an efficient search
strategy, several  measures can be used.  The most obvious  one is the
\textit{rate of success}, which of course involves a proper definition
of successful/failed searches. We shall consider a search to be failed
if the  search time exceeds an upper  bound, or if the  maximum of the
probability distribution  when the  entropy falls below  the detection
threshold does not coincide with the real position of the source.  The
next measure of performance is the mean search time, together with its
fluctuations.

The  motivation of  this work  is geared  towards applications  in the
development of  future sniffers and their use  for resolving practical
problems. The paper  is organized as follows: after  a brief review of
the infotaxis algorithm in Section~\ref{sec:infotaxis}, we discuss its
performance  as a  function of  the parameters  of the  environment in
Section~\ref{sec:param_analisis}.                                    In
Section~\ref{sec:misspecifications} we perform a quantitative analysis
of  the drop  in  performance due  to  an imperfect  modelling of  the
transport  process in the  environment. Finally,  a discussion  of the
results is presented in Section~\ref{sec:summary}.

\section{Infotaxis}
\label{sec:infotaxis}

In this  section we briefly  describe the infotaxis  search algorithm,
and  refer the  interested reader  to Ref.~\cite{Vergassola}  for more
details  and  insights  (see  also section  II  of  \cite{duque2014}).
Infotaxis was designed as an olfactory search strategy that is able to
find the location  of a target that is  emitting chemical molecules to
the environment  which is  assumed to be  turbulent \cite{Vergassola}.
By  decoding  the trace  of  detections  and  non detections  of  such
chemicals $\mathcal{T}_t$,  the infotactic searcher  solves a Bayesian
inference problem to reconstruct at  each time a probabilistic map for
the  position of  the  target. This  map,  commonly named  \emph{belief
  function} in the  context of information theory, is  refined in time
by the searcher  by choosing its movements as  those that maximize the
local gain of information. A  suitable indicator of a successful search
is the Shannon entropy  associated to the belief function, approaching
zero when the belief function  becomes a delta function located at the
position of the target.

The infotaxis search  strategy has two key elements:  on one hand the
average  rate of detections  $R(\mathbf{r},\mathbf{r}_0)$, which  is a
function  of  the searcher's  position  $\mathbf{r}$  and the  assumed
target's position $\mathbf{r}_0$, and on the other hand the belief
function itself $P_t(\mathbf{r}_0)$. 

The  rate function  $R$  models how  the  the chemicals  emitted at  a
position $\mathbf{r}_0$ are transported  by the environment, and it is
usually taken to be the solution of an advection-diffusion equation in
free  space \cite{Vergassola}.   In two  dimensions,  the rate
function becomes
\begin{equation}
R(\mathbf{r},\mathbf{r}_0)= \frac{\gamma}{\ln \left( \frac{\lambda}{a} \right)} e^{\frac{V\left( y_0 - y \right)}{2D}} K_0 \left( \frac{|\mathbf{r} - \mathbf{r}_0|}{\lambda} \right) \ ,
\end{equation}
where  $\gamma$ is the  rate of  emission of  chemicals, $D$  is their
isotropic effective diffusivity, $a$ is the characteristic size of the
searcher,  $K_0$ is  the  modified  Bessel function  of  order 0,  and
$\lambda$ the \textit{correlation length}, given by
\begin{equation}\label{eq:lambda}
 \lambda = \left( \frac{D \eta}{1 + \frac{V^2 \eta}{4D}} \right)^{1/2}
\end{equation}
where $\eta$  is the  lifetime of the  emitted molecules, and  $V$ the
mean current or wind (which  blows, without loss of generality, in the
negative  $y$-direction).  The  correlation  length $\lambda$  can  be
interpreted  as the  mean  distance traveled  by  a volatile  particle
before it decays.

The  rate  function  is  used  by  the  Bayesian  inference  analysis,
weighting the  actual number of  detections with the expected  one, to
reconstruct the belief  function representing the searcher's knowledge
about the target's location.  This function is a time-varying quantity
that is updated, given the trace of detections $\mathcal{T}_t$ at time
$t$, using the Bayes' formula. If one assumes statistical independence
  of successive detections  (i.e. a Poisson process)  the probability
function at time $t$ posterior to experiencing a trace $\mathcal{T}_t$
is given by:
\begin{equation}
 P_t(\ro) = \frac{\mathcal{L}_{\ro}(\mathcal{T}_t)}{\int \mathcal{L}_{x}(\mathcal{T}_t) dx} \ ,
\end{equation}
where    
\[\mathcal{L}_{\ro}     =    e^{-\int_0^t    R(\r(t')|\ro)dt'}
\prod_{i=1}^H  R(\r(t_i)|\ro)\]   and  $H$  is  the   total  number  of
detections   registered   by   the   searcher  at   successive   times
$(t_1,\dots,t_H)$.

The  searcher uses  the belief  function, choosing  its  movements not
towards  the most  probable value  of $P_t(\mathbf{r}_0)$  but  to the
position at which  the expected gain of information  about the target's
position is  maximized.  Assuming that  the search domain is  a square
lattice  and quantifying  the uncertainty  of the  searcher  about the
target's   position   with   the   Shannon   entropy   associated   to
$P_t(\mathbf{r}_0)$, the maximization  process means that the searcher
moves  from  its  current  position  $\mathbf{r}$ at  time  $t$  to  a
neighboring position $\mathbf{r'}$  at time $t + \delta  t$, for which
the  decrease in  entropy  $\overline{\Delta S}(\r\rightarrow\r')$  is
largest.

The expected variation of entropy upon moving from $\r$ to $\r'$ is given by
\begin{equation}\label{deltaS}
\overline{\Delta S}(\r\rightarrow\r') = -P_t(\r^\prime) S +
(1 - P_t(\r'))\left[\sum_{k=0}^\infty \rho_k(\r') \Delta S_k \right]
\end{equation}
where \[\rho_k(\r')  = h(\r')^k  e^{-h(\r')}/k!\] is the  probability of
having  $k$ detections  during the  time  $\delta t$,  with \[h(\r')  =
\delta t \int P_t(\ro) R(\r'|\ro)  d\ro\] the mean number of detections
at  position $\r'$,  and $\Delta  S_k$  is the  expected reduction  in
entropy assuming  that there  will be $k$  detections during  the next
movement. The  first and  second  term in  Eq.  (\ref{deltaS})  evaluate
respectively the reduction in entropy if the target is found or not at
$\mathbf{r'}$ in the next  step. Therefore, Eq. (\ref{deltaS}) naturally
represents a balance between exploitation and exploration.

The numerical experiments  reported in the rest of  this paper are set
as  follows:  At  time  $t=0$  the  search  starts  with  a  uniformly
distributed  belief  function, \emph{i.e.},  the  searcher is  totally
ignorant about  the target's position. The initial  state is therefore
of maximal entropy.  The search  ends when the Shannon entropy takes a
value  below a  certain threshold,  which we  set to  $S_\varepsilon =
10^{-4}$  (first  hitting  time  criterion).  During  the  search  the
associated entropy  approaches zero, not  necessarily monotonously, as
the belief function gets narrower and under very general circumstances
it becomes  a delta peak centered  at the target's  location.  We will
show however that this may not always be  the case.  This  motivates us to distinguish two
different  situations for  an  unsuccessful search:  when the  entropy
threshold is  reached but  the maximum of  the belief function  does not
coincide with the position of the source  (type I), and  when the search  exceeds the
maximum time limit $T$ without reaching  the entropy threshold
(type II). 

\section{Dependence on the environment parameters}
\label{sec:param_analisis}

We  first study the  dependence of  the search  time on  the different
parameters  involved in  the environment  model, namely  the diffusion
coefficient $D$ determining the typical  size of the area the searcher agent
explores  between  successive  updates  of the  belief  function,  the
emission  rate  $\gamma$ related  to  the  amount  of information  the
searcher can  receive through  the detections and  the wind  speed $V$
that breaks the  symmetry of the search by  distinguishing the regions
of  the search  domain where  the target  is most  likely  located. We
recall  that changes  in $D$  and  $V$ modify  the correlation  length
Eq.~\ref{eq:lambda},  that  roughly speaking,  determines  the way  in
which the searcher approaches the target.  Naturally, the modification
of any  of these  parameters is reflected  on the balance  between the
explorative     and    exploitative     tendencies     of    infotaxis
\cite{Vergassola}.

To  be  precise,  we  consider   a  search  domain  consisting  on  a
two-dimensional  lattice   of  size  $100\times100$   with  reflecting
boundary  conditions, meaning  that if  at any  instant the  agent is
located on  the boundary  of the search  domain the  movement pointing
outward is  supressed. In the  numerical experiments reported  in this
section,  the  target  is  located  at coordinates  $(0,35)$  and  the
searcher is  placed initially at  $(0,-47)$.  All positions  are given
with  respect  to  the central  lattice site  $(0,0)$ of  the search
domain.  Furthermore, we  impose that the search starts  at time $t=0$
with  the searcher  having registered one detection.  The  size of  the
searcher is  set to $a =  1$ and the  molecule's life time to  $\eta =
2500$.

\subsubsection{Diffusion coefficient}

\begin{figure}[!t]
  \centerline{\includegraphics*[width=0.45\textwidth]{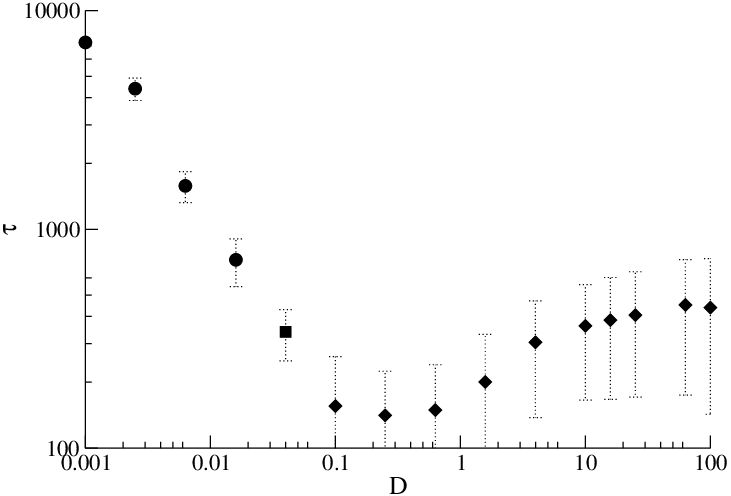}}
  \caption{Search time as a function of the diffusion coefficient $D$,
    obtained as  an average  over $1000$ trajectories.   The different
    symbols identify  the direction of  the initial step  the searcher
    takes  at time $t=1$:  circle (down),  square (left/right)  and diamond
    (up).   The  error  bars   correspond  to  the  data's  standard
    deviation. The rest of the parameters were set to $a = 1$, $\gamma
    = 1$, $\eta = 2500$ and $V=0$.}
  \label{fig:D}
\end{figure}

In this section we study the  variation of the search time with $D$ in
the absence  of wind $V=0$. Note  that with this choice  any change in
$D$  corresponds  to a  quadratic  change  in  the correlation  length
$\lambda$  (see  Eq.~\ref{eq:lambda}).   These  results are  shown  in
Fig.~\ref{fig:D}, where  we can distinguish two  different regimes: At
small diffusivities the search  time decreases two orders of magnitude
as  $1/D$, reaching  a minimum  value at  $D \approx  0.5$.  At larger
diffusivities $\tau$ increases and saturates at $\tau \approx 400$.

Both regimes can be understood simply in terms of the variation of the
correlation length  $\lambda$. At small  diffusivities the correlation
length  is small,  meaning that  the effective  area inside  which the
Bayesian  inference  has an  effect  is  small  compared to  the  whole
domain. As  $D$ increases $\lambda$  increases and the  search becomes
more effective  as this  implies an increase  of the  effective region
where the  searcher explores to find the  source's position, enhancing
the  searcher's ``field of vision''.   For $D\approx0.5$  the correlation
length  becomes  of  the  same  order  of the  length  of  the  domain
($\lambda\approx50$), and the minimum search time is attained.  At this
point the  exploitative terms in  infotaxis become important.   At the
second  regime where the  correlation length  becomes larger  than the
search domain the infotactic search looses resolution as larger values
of $\lambda$  entail further  uncertainty about the  source position,
and the search time increases  again and saturates. In the presence of
wind $V\ne0$, the same qualitative behaviour is expected.

It is interesting to note that the fluctuations around the search time
also have a different behaviour in these two regimes. The behaviour of
the  fluctuations  was   recently  studied  in  \cite{duque2014},  and
associated  to  the  direction  of  the  initial  step  taken  by  the
searcher. There  it was  found that the  initial step in  infotaxis is
fully determined by the geometry of the boundary and by the searcher's
proximity to it, forming a  partition with elements of similar initial
behaviour. More importantly, the area and shape of the elements of the
partition  was  mainly  affected  by  the  value  of  the  correlation
length. Therefore, for  a fixed initial position of  the searcher, a
variation  in $\lambda$ might change  its initial  step and the  different
symbols  in Fig.~\ref{fig:D} distinguish  this initial  behaviour. The
increase of fluctuations around the  search time in the regime of
large  diffusivity  is in  agreement  with  our  previous findings  in
\cite{duque2014}.

\subsubsection{Wind velocity}

\begin{figure}[!t]
  \centerline{\includegraphics*[width=0.45\textwidth]{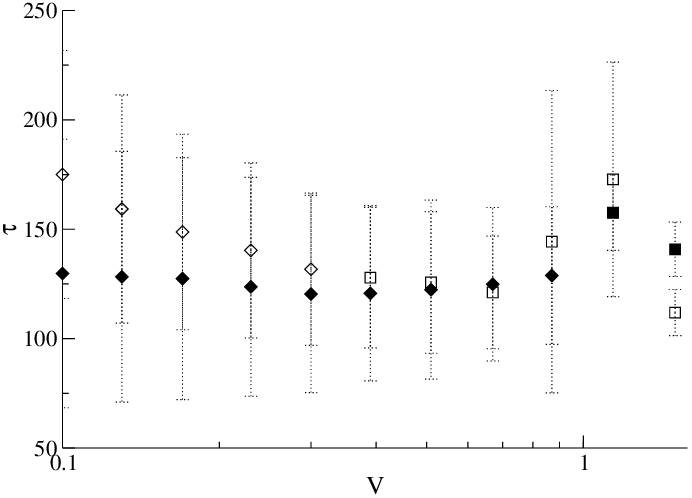}}
  \caption{Search time  as a  function of the  wind speed $V$  for two
    different  starting positions  of  the searcher:  $r_o =  (0,-47)$
    (solid  symbols)  $r_o =  (47,0)$  (empty  symbols).  The rest  of
    parameters were set to $a = 1$, $\eta = 2500$, $D = 1$ and $\gamma
    = 1$.  }
  \label{fig:V}
\end{figure}

We now turn our attention to  the dependence of the search time on the
wind speed  $V$. We  show this in  Fig.~\ref{fig:V} for  two different
starting  positions   of  the  searcher:   $(0,-47)$  (solid  symbols)
corresponding  to a searcher  starting inside  the region  of frequent
detections and $(47,0)$ (empty symbols)  at which the searcher is in a
region  of low  detections. The  presence  of wind  breaks the  radial
symmetry of  the search and more importantly,  changes the correlation
length $\lambda$. This will affect  not only the mean search time but
its fluctuations as discussed in \cite{duque2014}. However, the search
time does  not seem  to change  much with the  variation of  the wind
speed. Moreover, we observe that  the dependence of $\tau$ on the wind
speed  is  qualitatively  the  same  irrespectively  of  the  starting
position of the searcher.

\subsubsection{Emission rate} 
\label{Emission-rate}

\begin{figure}[!t]
  \centerline{\includegraphics*[width=0.45\textwidth]{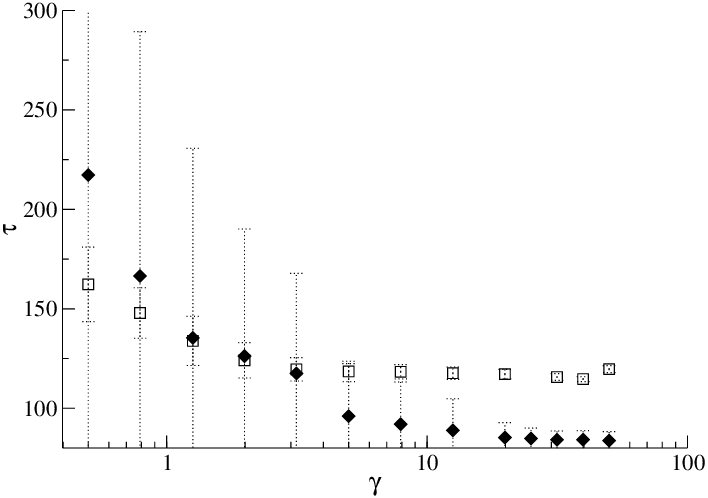}}
  \caption{Search time as a function  of emission rate $\gamma$ for $V
    = 0$  (solid symbols) and  $V=-1.5$ (empty symbols).  The  rest of
    the parameters were set to $a = 1$, $\eta = 2500$ and $D = 1$.}
  \label{fig:gamma}
\end{figure}

Larger emission rates mean that the source emits more information about
its  presence to  the environment,  which  in turns  implies that  the
searcher will  have more information about the  source. This
is  what we  observe  in Fig.\ref{fig:gamma},  where  the search  time
decreases with increasing emission rate $\gamma$, independently of the
magnitude of the  wind. Interestingly, we find that  at large emission
rates $\gamma\approx5$, the search last  less at zero wind than in the
presence of it. At first sight this appears counterintuitive since the
presence of wind acts as an additional source of information about the
direction in which  the source is located. However,  we have found that
these longer  search times in the  regime of large $\gamma$  are due to
the additional time the searcher spends during the initial explorative
zigzagging motion when it is far from the source and the detections
are scarce. In the absence of wind the searcher tends to move
directly to the center of the domain, thus closer to the source and to
the region in which the detections are more frequent.

We  finish  this section  discussing  the  evaluation  of the  entropy
variation  involved  in  each   of  the  possible  searcher  movements
(Eq.~\ref{deltaS}).   The  numerical  computation of  Eq.~\ref{deltaS}
requires to  truncate the infinite  sum corresponding to  the weighted
probability  of having any  possible number  of detections  during the
searcher motion from  $\r$ to $\r^\prime$.  We do  this by summing all
terms  until the cumulative  probability of  $k$ detections  reaches a
value close  to $1$  ($0.999$ in our  computations). However,  at high
emission   rates  the   mean  number   $k$  of   detections  increases
drastically, demanding  a much larger  number of terms to  consider in
the infinite sum, entailing an important increase of the computational
cost.   To  keep  the  infotaxis  computationally  efficient  we  have
approximated the  entropy variation of  Eq.~\ref{deltaS} by truncating
the infinite sum to a maximum number of detections $k_{\mathrm{max}}$,
irrespectively   of   the   value   of  the   cumulative   probability
$\sum_k\rho_k$, and  found some  interesting aspects of  the infotatic
search that we discuss now.

\begin{figure}[!t]
  \centerline{\includegraphics*[width=0.45\textwidth]{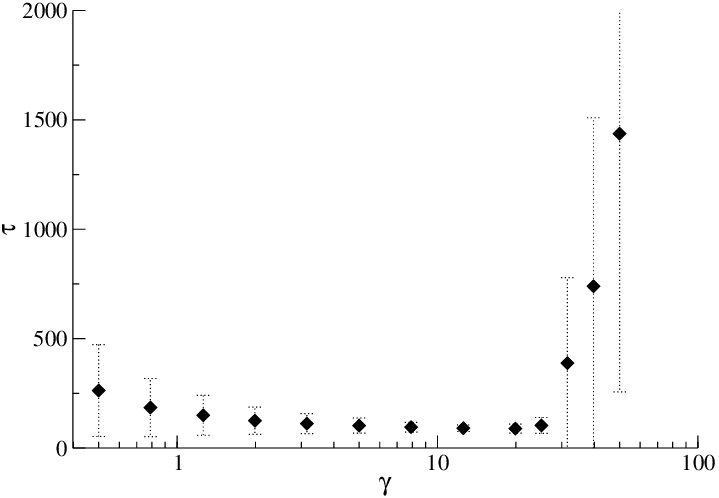}}
  \caption{Search time as a function of the emission rate $\gamma$ in
    the absence of wind $V = 0$ and for a truncated sum in Eq.~\ref{deltaS} with $k_{max} <= 20$. 
   The rest of the parameters were set to $a = 1$, $\eta = 2500$ and $D = 1$.}
  \label{fig:gamma_ii}
\end{figure}

In  Fig.~\ref{fig:gamma_ii} we show  the dependence  of $\tau$  on the
emission  rate in the  absence of  wind $V=0$,  truncating the  sum in
Eq.~\ref{deltaS}  to  $k_{\mathrm{max}}\le20$  and  the  rest  of  the
parameters as in Fig.~\ref{fig:gamma}.  Comparing these two figures we
observe that at  low emission rates both numerical  procedures lead to
the same results since the cumulative probability of $k$ detections is
one for  $k\le k_{\mathrm{max}}$.  At high emission  rates this  is no
longer  true.  Nevertheless  we  obtain that  the infotactic  searches
remain successful albeit with a much larger search time.

\begin{figure}[!b]
  \centerline{\includegraphics*[width=0.45\textwidth]{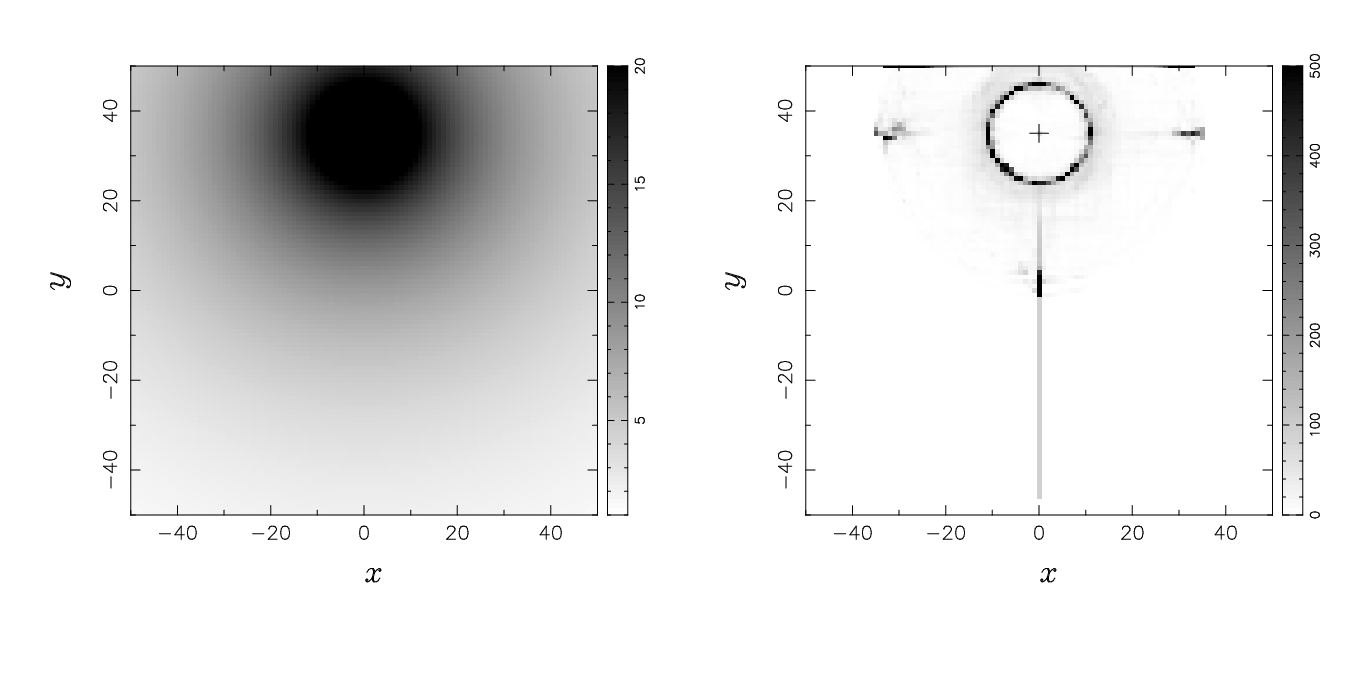}}
  \centerline{\includegraphics*[width=0.45\textwidth]{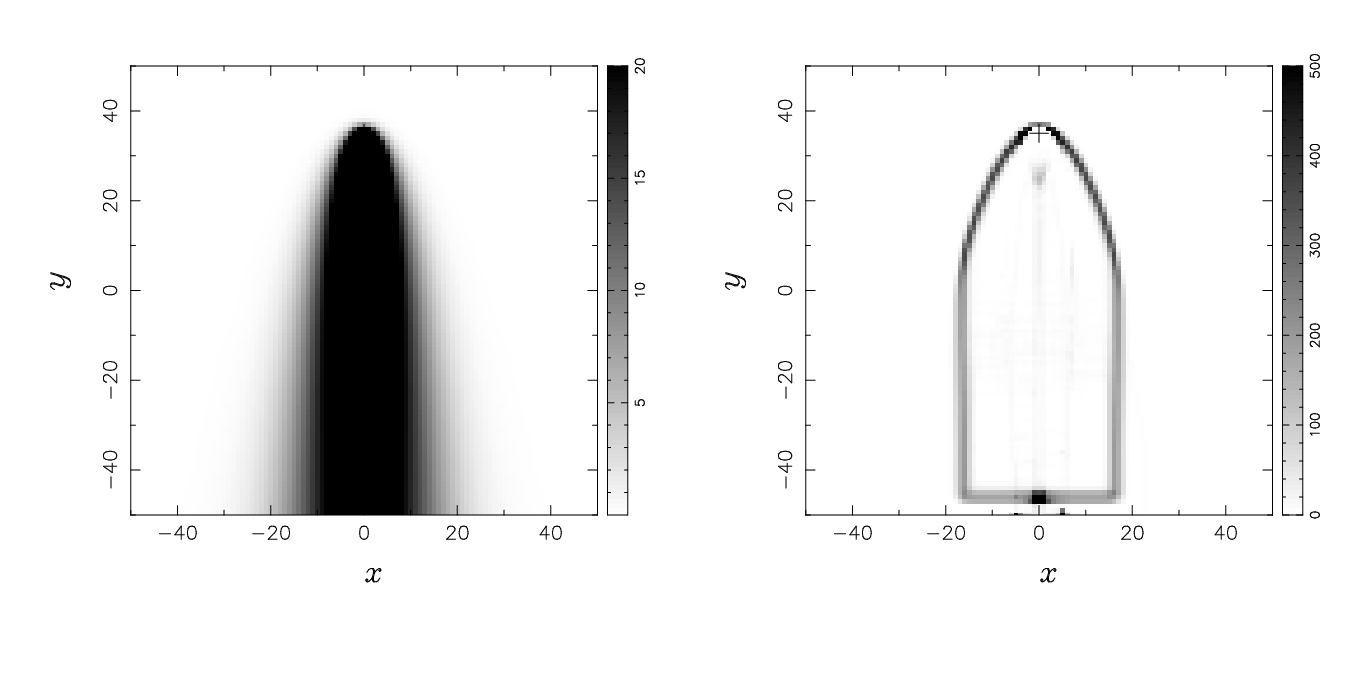}}
  \caption{Density of visited sites for successful searches for which
    the searcher does not reach the source. The left panels show a
    density map of the mean chemical concentration field. The right
    panels show the density of visited sites by 200 search
    trajectories, in the absence of wind $V=0$ (upper row) and with a
    wind of $V=1.5$ (lower row). The rest of the paremeters were set
    to $a = 1$, $\eta = 2500$, $D = 1$ and $\gamma = 50$.}
  \label{fig:Maps_gamma}
\end{figure}

To understand the consequences  of approximating the truncation of the
infinite  sum  we have  studied  the  global  topology of  the  search
trajectories.   In Fig.~\ref{fig:Maps_gamma}  we show  the  density of
visited sites of the trajectories  that lead to a successful search in
the   absence  (upper  row)   and  presence   (lower  row)   of  wind.
Surprinsingly,  under this  approximation we  observe that  the belief
function peaks  exactly at  the source even  thugh the  searcher never
reaches  the  source position  but  get stucks  away  of  it. This  is
evidenced  in the  density of  visited sites  in the  right  panels of
Fig.~\ref{fig:Maps_gamma}. As a matter of fact, we have found that the
searcher remains  for long times over the  density curve corresponding
to $R(r|r_0)  = k_{\mathrm{max}}$.  In  this region the agent  feels a
number of  detections that  would correspond to  be very close  to the
source, thus  changing from an  explorative search to  an exploitative
one,  emphasizing   a  major  contribution   of  the  first   term  of
Eq.\ref{deltaS} in the decision making processs.  This is evidenced by
comparing the highest density of  visited sites on the right column of
Fig.~\ref{fig:Maps_gamma}  with the  shape of  the  corresponding mean
concentration  field that  we  show in  the  left column  of the  same
figure.   Notwithstanding  this, the  Bayesian  inference continue  to
refine the  belief function by making  the probabilistic triangulation
from  a distance,  until it  becomes  a peaked  distribution over  the
source position.

This surprising effect  stresses one of the most  important sources of
the robustness of  the infotaxis search strategy: the  location of the
source is possible even if the searcher never reaches its position.

\section{Performance of infotaxis under innacurate modelling}
\label{sec:misspecifications}

In this Section we focus  on the performance of infotaxis, as measured
by the success  rate and mean search time, when  the searcher does not
have an exact knowledge of the parameters of the transport process. It
is natural to  expect a drop of perfomance in this  regime, but we are
interested in  a quantitative analysis.   It is hard  to overemphasize
how  important this  matter is  for practical  purposes,  as measuring
devices  introduce  some  uncertainty  in  the best  case,  and  other
parameters that are harder to measure can only be estimated.

\subsubsection{Mismatches in $\lambda$}

\begin{figure}[!t]
  \centerline{\includegraphics*[width=0.25\textwidth]{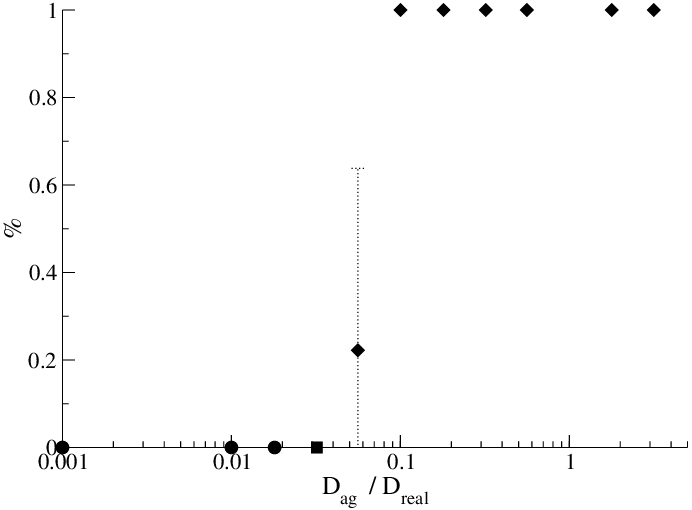}\includegraphics*[width=0.25\textwidth]{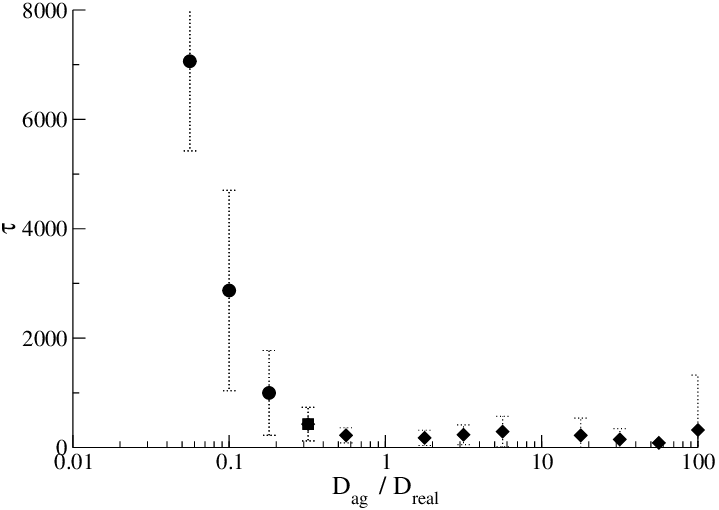}}
  \caption{Performance  as  a  function  of  inaccurate  modelling  of
    $\lambda$.  Left  panel: Success  rate. Right panel:  Search time.
    Rest  of parameters:  $a =  1$, $\eta  = 2500$,  $\gamma =  1$ and
    $D_{real}=1$.}
  \label{fig:fpt_D_est}
\end{figure}

We  begin our performance  analysis with  the misspecification  in the
correlation  length parameter $\lambda$,  Fig.\ref{fig:fpt_D_est}.  We
recall that $\lambda$ is defined in \eqref{eq:lambda}, so we will keep
the rest of the parameters  constant and let the diffusion coefficient
$D$ change.  We shall  denote by $D_{agent}$ the diffusion coefficient
used by  the searcher  for his Bayesian  inference and  $D_{real}$ the
true diffusivity of  the transport process (and likewise  for the rest
of the parameters).

Our   results   (see  Figure   \ref{fig:fpt_D_est})   show  that   for
$\lambda_{agent}>\lambda_{real}$  the   performance  of  infotaxis  is
largely  unaffected by  the  mismatch:  the sucess  rate  is close  to
$100\%$  and the  mean search  time is  close to  the case  of perfect
knowledge.   As  in the  previous  section,  when $\lambda_{agent} \gg 
\lambda_{real}$, the  searcher assumes that  the information collected
during the  search process comes from  a region larger  than it really
is, causing a slower learning to find the source position and thereby,
a larger  search time.  However, such  increase in the  search time is
hardly  observed in  this case  due to  the dilute  conditions  of the
search and the particular starting position of the searcher chosen for
the numerical  simulations (its  first step at  low detection  rate is
persistently directed towards the source).

However,  an underestimation  of $\lambda$  causes a  drastic  drop in
performance. This is specially  evident when $\lambda_{agent}$ is less
than  or of the  order of  the initial  distance of  the agent  to the
boundary of  the search domain.  In  these cases, the  initial step of
the search changes and the  search time increases because the searcher
explores  the space  and learns  about  the source  position in  steps
smaller than  it should.  We  should remark that all  the unsuccessful
searches occur when $\lambda$ is underestimated and they correspond to
type I failures:  the maximum of the belief  function when the entropy
threshold is reached  does not coincide with the  true position of the
source.

\subsubsection{Mismatches in $\gamma$}

Perhaps the most interesting parameter  to analyze is the rate of odor
emission  $\gamma$.  It  is  worth  stressing  that  while  the  other
parameters of  the transport process,  such as the diffusivity  an the
wind velocity can be measured  with appropriate equipment, the rate of
emission of volatile  particles that are transported by  the medium is
harder  to  measure  and  subject  to  greater  variability,  e.g.  if
infotaxis is used by a robotic agent to find the source of a plague in
a  crop field,  the  emission rate  of  volatiles will  depend on  the
biological state of the infected plant \cite{vanHenten2011,pisa}.

\begin{figure}[!t]
  \centerline{\includegraphics*[width=0.45\textwidth]{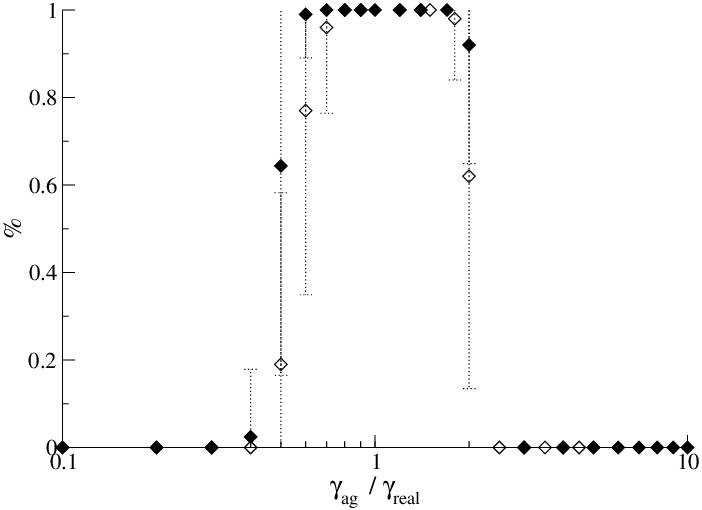}}
  \caption{Success   rate.   Black:   $\gamma_{real}  =   1$.   White:
    $\gamma_{real} = 10$.  Rest of parameters: $a=1$, $\eta = 2500$ and $D_{real} = D_{agent}  = 1$.}
  \label{fig:Success_gamma_est}
\end{figure}

\begin{figure}[!t]
  \centerline{\includegraphics*[width=0.25\textwidth]{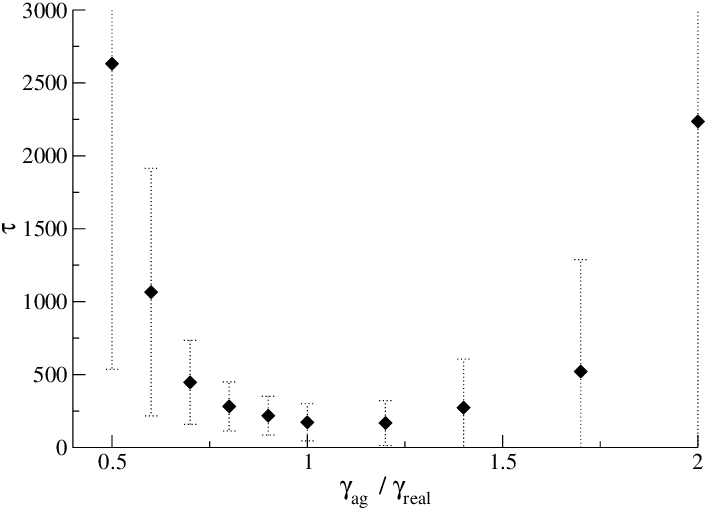}\includegraphics*[width=0.25\textwidth]{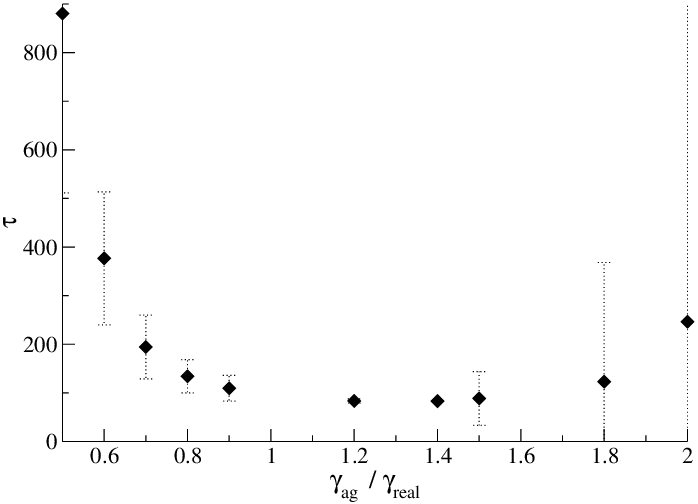}}
  \caption{Search time for the successful searches. Left panel: $\gamma_{real} = 1$. Right panel: $\gamma_{real} = 10$. 
   Rest of parameters: Rest of parameters: $a=1$, $\eta = 2500$ and $D_{real} = D_{agent}  = 1$.}
  \label{fig:fpt_gamma_est}
\end{figure}

Figures  \ref{fig:Success_gamma_est} and  \ref{fig:fpt_gamma_est} show
the success rate and the variation of the search time as a function of
the mismatch in $\gamma$ for  two different emission regimes (i.e. two
different values of $\gamma_{real}$).

The      first     clear      observation     when      looking     at
Fig.~\ref{fig:Success_gamma_est}  is that, as  opposed to  the results
exhibited  in  Section  \ref{sec:param_analisis},  the search  is  not
always  successful.    Indeed,  there  is   a  window  of   values  of
$\gamma_{agent}$    centered     around    the    perfect    knowledge
($\gamma_{agent}=\gamma_{real}$)    where     infotaxis    is    still
feasible. This window corresponds  to the interval $\gamma_{agent}/\gamma_{real} \in
[0.5,2.5]$ and  seems to be independent of  the value of
$\gamma_{real}$.

On both sides of the window of admissible estimated values of $\gamma$
the  search fails for  two different  reasons.  An  underestimation of
$\gamma$  ($\gamma_{agent}<\gamma_{real}$) leads  to type  I failures,
while an  overestimation of $\gamma$  leads to type II  failures.  The
mean search time  of the successful searches reaches  a minimum in the
case of perfect knowledge and grows  on both sides, as shown in Figure
\ref{fig:fpt_gamma_est}     for     two     different    values     of
$\gamma_{real}$. The reason of this  increase in mean search time is a
deficient Bayesian inference,  as the agent believes the  source to be
farther away  or much  closer than it  really is  for a given  rate of
detection.  In other words,  the Bayesian  inference converts  a given
rate of detection to a given  distance of the agent to the source, and
this conversion is  not accurate when the estimated  value of $\gamma$
in use by the agent differs from the real one.

\begin{figure}[!h]
  \centerline{\includegraphics*[width=0.25\textwidth]{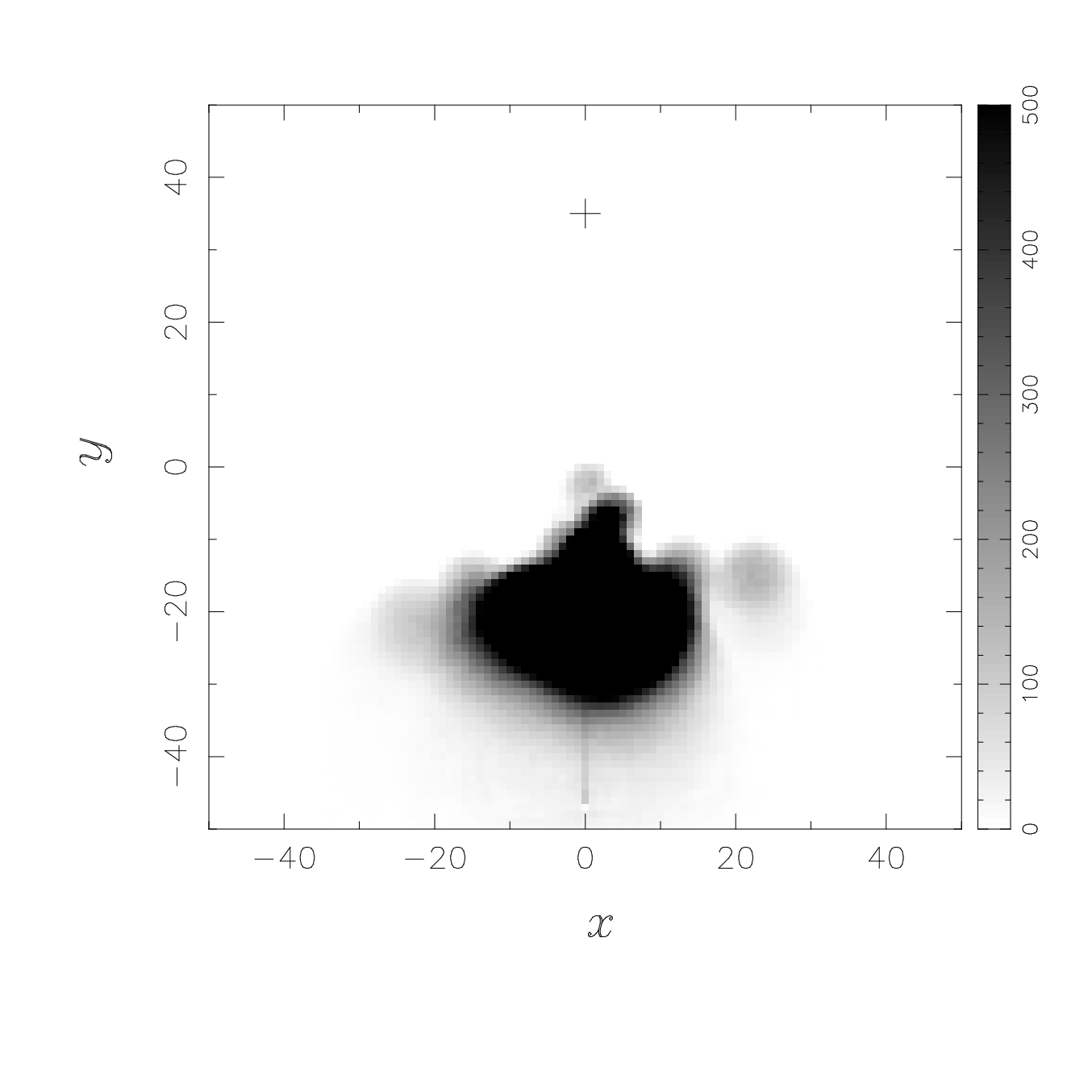}\includegraphics*[width=0.25\textwidth]{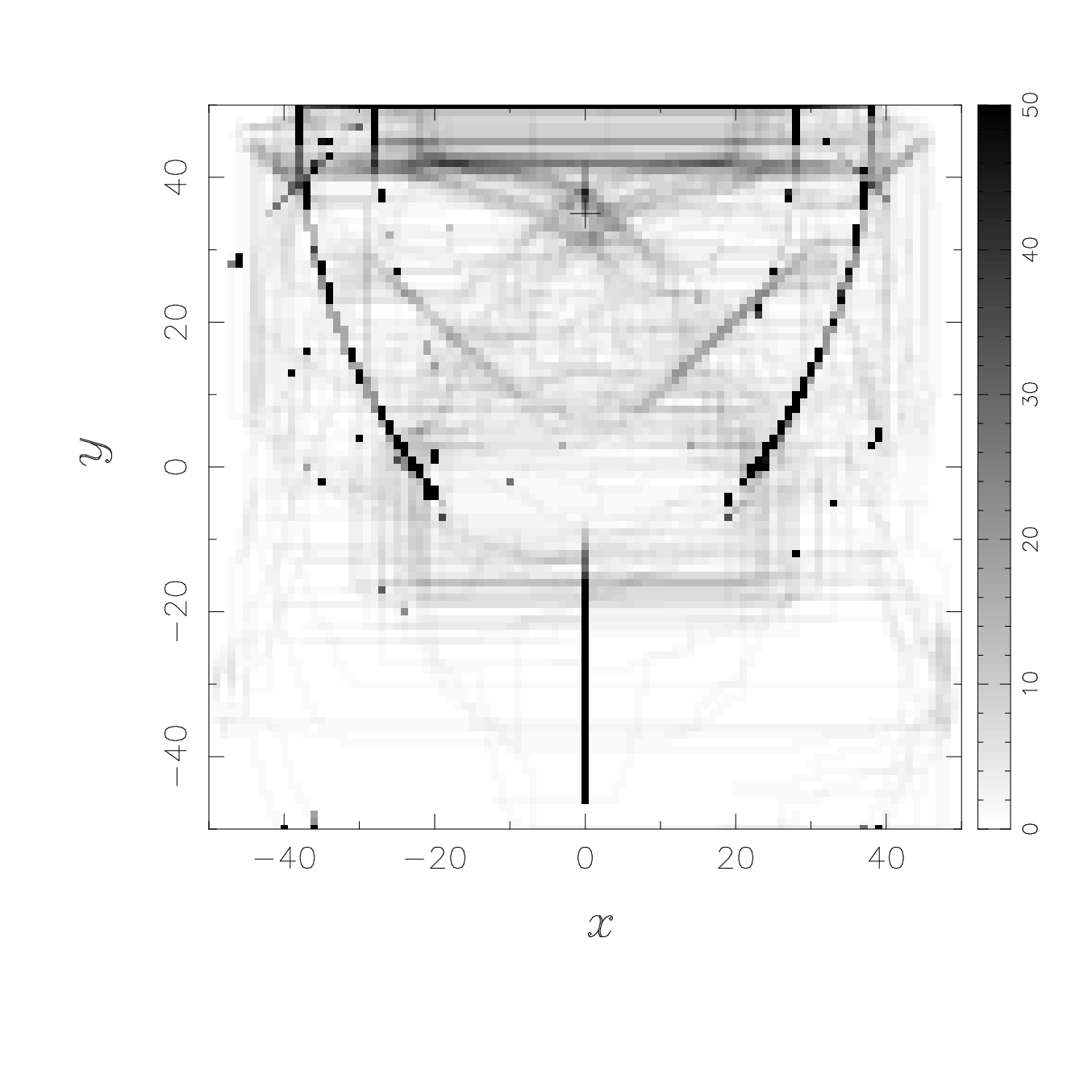}}
  \caption{Unsuccessful searches. Left panel: Underestimation of the emission rate: $\frac{\gamma_{agent}}{\gamma_{real}} = 0.1$. 
   Right panel: Overestimation of the emission rate: $\frac{\gamma_{agent}}{\gamma_{real}} = 3.5$ 
   Rest of parameters: Rest of parameters: $a=1$, $\eta = 2500$ and $D_{real} = D_{agent}  = 1$.}
  \label{fig:Maps_gamma_est}
\end{figure}

We  have explored  in greater  detail the  patterns of  motion  of the
searcher under the  two situations, underestimation and overestimation
of $\gamma$.   In Figure~\ref{fig:Maps_gamma_est} we  plot the density
of sites visited by the searcher in 100 trajectories starting from the
same initial  position with the source  in the same  position.  In the
left  panel, corresponding  to  $\gamma_{agent}=0.1\gamma_{agent}$ the
agent spends  most of the time  exploring its vicinity  well away from
the source. It shows a random  motion similar to the final steps of an
infotactic search  in the vicinity  of the source.  Underestimation of
$\gamma$ causes  the agent to believe  that the source  is much closer
than it really is.

On the  other hand, when  $\gamma_{real}$ is overestimated,  the agent
believes the  source to be  farther than it  is, and often  the belief
function  concentrates on the  domain boundary,  specially in  the top
corners. This explains the density of sites visited by the searcher in
the right panel of Fig. 9, where most of the trajectories involve more
deterministic and  persistent motions,  as in the  initial steps  of a
normal search when  the agent is far away from the  source. Due to the
symmetry  of the problem,  the belief  function concentrates  for some
time in  one corner,  but then it  shifts to  the other corner  as the
searcher approaches it and discovers that the source is not there. The
searcher enters into a loop that  ends up in a frozen position, due to
an  effect similar to  the one  described in  Section III.3,  which is
caused by  an underestimation of  $k$ (the agent registers  much fewer
detections than the number it  expects from its belief function). As a
result, the  search terminates  in a type  II failure, as  the maximum
time  is  reached  before   the  entropy  falls  below  the  detection
threshold.

\section{Conclusions}
\label{sec:summary}

We have studied the performance  of the infotaxis search strategy as a
function of  the parameters  of the transport  process as well  as its
performance   with  respect   to   an  inaccurate   modeling  of   the
environment. We  have assessed these  questions by means  of intensive
numerical simulations, and  we have shown the variation  of the search
time and the success rate of infotaxis in all the different cases.  In
our implementation of infotaxis we use the first hit as opposed to the
first  passage criterion,  i.e.  vicinity in  information rather  than
physical space.

We have  shown, in accordance  with the previous literature,  that the
search time  shows strong dependence not  only of the  initial step of
the search,  but also in  the way in  which the searcher  explores the
environment  (mainly determined by  the correlation  length $\lambda$)
and exploits the information  collected during the search process.  In
the case of  a perfect knowledge of the environment,  we find that the
searches are always  successful but the mean search  time changes with
the  parameters  of the  transport  process.   As  a function  of  the
correlation length  $\lambda$, the mean search time  reaches a minimum
value  when  $\lambda$  has  the  size  of  the  search  domain.   The
dependence of the  mean search time of the wind  velocity is very mild
as  well  as its  dependence  on the  initial  position  of the  agent
relative  to the  source and  wind  direction.  The  mean search  time
decreases with  emission rate $\gamma$, as information  is released to
the  agent at  a  higher rate.   However,  at very  high $\gamma$  the
computational complexity of the algorithm increases, and we have found
that  simplifying the  computation still  leads to  succesful searches
even when the agent never reaches the source.

We  have studied the  drop in  performance of  infotaxis caused  by an
imperfect modelling of the environment expressed through an inaccurate
estimation  of the parameters  of the  transport process.  Our results
show  that  in  practical  cases  it  is  safer  to  overestimate  the
correlation  length $\lambda$  than  to underestimate  it,  as in  the
former case  no significant  drop in performance  occurs while  in the
latter the sucess rate quickly drops.  The situation is different when
the mismatch between real and  estimated value occurs for the emission
rate $\gamma$.  In this case there  is a window around  the real value
where infotaxis remains  robust, but overestimation or underestimation
by a factor  of two leads to a rapid decay  in performance, with lower
success rates and higher mean search times.

Our results  places some limits  on the performance of  infotaxis, and
have practical  consequences for the design of  future infotaxis based
machines  to track  and  detect  an emitting  source  of chemicals  or
volatile substances.


\acknowledgments This work  has been supported by Grant  No. 245986 of
the  EU project  Robots  Fleets for  Highly  Agriculture and  Forestry
Management.   J.D.R.   was  also  supported by  a  PICATA  predoctoral
fellowship   of  the  Moncloa   Campus  of   International  Excellence
(UCM-UPM).   The research  of D.G.U.   has been  supported in  part by
Spanish     MINECO-FEDER     Grants     No.      MTM2012-31714     and
No. FIS2012-38949-C03-01.   We acknowledge the use of  the UPC Applied
Math     cluster     system     for    research     computing     (see
http://www.ma1.upc.edu/eixam/index.html).  CMM has  been  supported by
the Spanish MICINN grant MTM2012-39101-C02-01.

\end{document}